\documentstyle[twocolumn,aps,epsfig]{revtex}
\def\be{\begin{equation}}
\def\e#1{\label{#1}\end{equation}}
\def\bea{\begin{eqnarray}}
\def\ea#1{\label{#1}\end{eqnarray}}
\def\r#1{(\ref{#1})}
\def\ee{\end{equation}}
\def\eea{\end{eqnarray}}
\def\bem#1{\begin{mathletters}\label{#1}}
\def\eml{\end{mathletters}}

\begin{document}
\draft
\title{
Relaxation of a two-level system strongly coupled to a reservoir:
Anomalously slow decoherence 
}
\author{A. G. Kofman}
\address{Chemical Physics Department, Weizmann Institute of Science, 
Rehovot 76100, Israel}
\date{May 17, 2001}
\maketitle
\begin{abstract}
Relaxation of a two-level system (TLS) into a resonant 
infinite-temperature reservoir with a Lorentzian spectrum is studied.
The reservoir is described by a complex Gaussian-Markovian field
coupled to the nondiagonal elements of the TLS Hamiltonian.
The theory can be relevant for electromagnetic interactions in 
microwave high-$Q$ cavities and muon spin depolarization.
Analytical results are obtained for the strong-coupling regime,
$\Omega_0\gg\nu$, where $\Omega_0$ is the rms coupling amplitude (Rabi
frequency) and $\nu$ is the width of the reservoir spectrum.
In this regime, the population difference and half of the initial 
coherence decay with two characteristic rates: the most part of the
decay occurs at $t\sim\Omega_0^{-1}$, the relaxation being reversible
for $t\ll(\Omega_0^2\nu)^{-1/3}$, whereas for 
$t\gg(\Omega_0^2\nu)^{-1/3}$ the relaxation becomes irreversible and
is practically over.
The other half of the coherence decays with the rate on the order of 
$\nu$, which may be slower by orders of magnitude than the time scale 
of the population relaxation.
The above features are explained by the fact that at $t\ll\nu^{-1}$
the reservoir temporal fluctuations are effectively one-dimensional 
(adiabatic).
Moreover, we identify the pointer basis, in which the reduction of 
the state vector occurs.
The pointer states are
correlated with the reservoir, being dependent on the reservoir phase.
\end{abstract}

\section{Introduction}

Decoherence of quantum systems is a subject of a significant
current interest \cite{ana00,zur00,dal01,fol}, since this phenomenon 
is of a great
importance both for the fundamentals of the quantum theory 
\cite{zur93} and for the field of quantum computation \cite{gru99}.
In most treatments of the decoherence and dissipation the thermal 
reservoir have been assumed to have a sufficiently broad 
spectrum (a short correlation time), so that the system-reservoir 
coupling can be considered weak, which allows one to use standard 
approaches, such as master equations \cite{ana00,sli78,coh92}.

For the generic case of a two-level system (TLS) this approach 
yields the familiar relation between the population (or longitudinal) 
and coherence (or transverse) relaxation times, $T_1$ and $T_2$
respectively: $T_1\ge T_2/2$.
The equality here is obtained in the case of a transverse reservoir, 
i.e., when the reservoir variables enter only the nondiagonal 
elements of the TLS Hamiltonian.
Otherwise, when the reservoir is coupled also to diagonal
elements of the Hamiltonian, the decoherence may occur much faster 
than the population relaxation.

Owing to the fast experimental progress in recent years, 
there appeared a number of novel materials and devices, such as
photonic-band structures \cite{yab87}, high-$Q$ cavities 
\cite{rai94,har94,kim94}, and
semiconductor heterostructures, many of which are of 
potential relevance for quantum information processing.
By creating structures in the electromagnetic continuum, such devices 
often produce a strong system-reservoir coupling.
In such cases the standard master-equation technique is inapplicable
and one should resort to other approaches.
Fortunately, for zero-temperature reservoirs, the TLS relaxation can 
be calculated for the general case in quadratures \cite{lou90}.
However, the finite-temperature case, which is important in the radio
and microwave frequency ranges, is much less understood.
Even the infinite-temperature limit, when the reservoir can be often
modeled by a classical random field, is insufficiently studied.

In this paper we report an analytical solution for the dynamics of 
a TLS which is strongly coupled to a transverse reservoir with a
Lorentzian spectrum in the infinite-temperature limit.
In this case the reservoir can be described by a classical field,
which is a complex Gaussian-Markovian random process 
\cite{cir91,uch00}.
This problem can be of relevance for microwave high-$Q$ cavities 
\cite{rai94,har94} at temperatures of several degrees Kelvin and 
higher.
It describes also a relaxation of a two-level atom in a resonant
chaotic field \cite{bur76,geo79,bur96}

Moreover, the above problem attracted a significant interest recently
\cite{uch00,shi87,sev89} in
connection with NMR and muon spin depolarization experiments.
In this case, numerical calculations \cite{sev89} showed
that the squared coherence may decay slower than the
population difference, violating thus the above inequality.
However, the above numerical studies did not elucidate the dependence 
of the relaxation on the parameters of the problem, not to mention a
physical explanation of the phenomenon.

Below we obtain a comprehensive picture of the TLS relaxation and
provide a physical interpretation of it.
In particular, we show that in the strong-coupling regime about half 
of the coherence still survives 
after the population relaxation is practically over!
The results obtained here are applied to the discussion of 
the pointer states, a concept of a significant current interest in
the theory of quantum measurements \cite{dal01,fol,zur81,zur93a}.

The paper is organized as follows.
In Sec. \ref{II} we formulate the problem and introduce an 
analogy between the Liouville equation for a TLS and the 
Schr\"{o}dinger equation for the spin 1.
In Sec. \ref{III} we discuss stochastic differential
equations for partial averages.
In Sec. \ref{IV} we present several limiting cases.
The population and coherence relaxation is investigated in Secs. 
\ref{VI} and \ref{VII}, respectively, and discussed in Sec. 
\ref{VIII}.
In Sec. \ref{IX} we obtain the pointer states and show that, in
contrast to the previously reported results, they are
correlated to the reservoir state.
Section \ref{X} provides concluding remarks.
The three Appendixes show details of the derivations.

\section{Formulation of the problem}
\label{II}

We consider a TLS with the states $|1\rangle$ and 
$|2\rangle$ and the resonant frequency $\Delta_0$.
The TLS Hamiltonian is
\be
H(t)=\hbar\Delta_0|2\rangle\langle 2|-
(\hbar/2)[\Omega_c(t)|2\rangle\langle 1|+\mbox{H.c.}].
\e{35}
The interaction of the TLS with a high-temperature reservoir is 
described by a complex function $\Omega_c(t)=u(t)+iv(t)$, which
is supposed to be a complex 
Gaussian-Markovian random process with the correlation function 
\be
\langle \Omega_c(t)\Omega_c^*(0)\rangle=\Omega_0^2e^{-\nu t},
\e{36}
where $\Omega_0$ is the rms coupling amplitude and $\nu^{-1}$ is the
correlation time of $\Omega_c(t)$.
The reservoir spectrum has a Lorentzian shape with the half width at
half maximum equal to $\nu$.

The Hamiltonian \r{35} has been used to describe muon spin 
depolarization experiments \cite{shi87,sev89}.
This Hamiltonian describes also [in the rotating wave
approximation (RWA) and in the interaction representation] a TLS with 
the resonance frequency $\omega_0$ coupled to a Lorentzian 
reservoir with the spectrum centered at $\omega_c$ (which is the case,
e.g., for microwave high-$Q$ cavities \cite{rai94,har94})
or to a laser chaotic field with 
the nominal frequency $\omega_c$ \cite{bur76,geo79,bur96}.
The RWA validity conditions are
$\Omega_0,\nu,|\Delta_0|\ll\omega_0$, where now 
$\Delta_0=\omega_0-\omega_c$.

It is convenient to write the TLS density matrix $\rho(t)$ as the 
column vector,
\be
r\equiv(r_1,r_0,r_{-1})^T=(\rho_{12},n,\rho_{21})^T,
\e{32} 
where $n=\rho_{11}-\rho_{22}$ is the population difference and
the superscript $T$ denotes the transpose.
Then the Liouville equation for $\rho(t)$ can be written in the form
\be
\dot{r}=A(t)r,
\e{63}
where
\be
A(t)=i\left(\begin{array}{ccc}
\Delta_0&-\Omega_c^*(t)/2&0\\
-\Omega_c(t)&0&\Omega_c^*(t)\\
0&\Omega_c(t)/2&-\Delta_0\end{array}\right).
\e{64}
The solution of Eq. \r{63} is $r(t)=G(t)r(0)$.
Here $G(t)$ is the Green function which obeys the equation
\be
\dot{G}=A(t)G
\e{65}
with the initial condition $G(0)=I$, where $I$ is the unity matrix.

It is well known that the Liouville equation \r{63}
can be cast as the equation \cite{fey57} 
\be
\dot{\vec{s}}=\vec{s}\times\vec{B}(t)
\e{38}
for the motion of a classical magnetic moment (pseudospin) 
\be
\vec{s}\equiv(s_x,s_y,s_z)=(2\text{Re}\rho_{21},
2\text{Im}\rho_{21},n)
\e{39}
in the effective magnetic field
\be
\vec{B}(t)=(u(t),v(t),\Delta_0).
\e{27}
This analogy helps to obtain an insight into the TLS behavior.

Moreover, it is of interest to consider another analogy, as follows.
Under the linear transformation $\psi=Sr$, where $S$ is a diagonal 
matrix with $S_{11}=-\sqrt{2},\ S_{00}=1$, and $S_{-1,-1}=\sqrt{2}$,
Eq. \r{63} becomes the {\em 
Schr\"{o}dinger equation} for a spin 1 in the magnetic field \r{27},
\be
\dot{\psi}=i\vec{B}(t)\cdot\vec{S}\psi.
\e{A1}
Here $S_i$ $(i=x,y,z)$ is the operator of the $i$th component of spin 
1 in the representation of the eigenfunctions of $S_z$ (the 
cyclic-basis representation) \cite{var88}.
[Equations \r{38} and \r{A1} imply that the effective gyromagnetic 
ratio equals 1.]
Note that the pseudospin $\vec{s}$ has the meaning of the polarization
vector \cite{var88} for the above spin 1.
The analogy \r{A1} can simplify calculations, by allowing one to use
the standard textbook techniques developed for the Schr\"{o}dinger 
equation.

\section{Equations for partial averages}
\label{III}

Consider the TLS density matrix averaged over such
realizations of the random process $\Omega_c(t)$ that assume the value
\be
\Omega_c=u+iv\equiv\Omega e^{i\phi}
\e{48}
at $t$.
This partially averaged density matrix $\rho(\vec{\Omega},t)$ written
as $r(\vec{\Omega},t)$ [see Eq. \r{32}], where $\vec{\Omega}=(u,v)$,
is given by
\be
r(\vec{\Omega},t)=G(\vec{\Omega},t)r(0).
\e{B1}
Here $G(\vec{\Omega},t)$, the partial average of $G(t)$,
obeys the equation \cite{bur76}
\be
\dot{G}=A(\vec{\Omega})G+LG, 
\e{47}
where $A(\vec{\Omega})$ is given by Eq. \r{64} with a constant
$\Omega_c$.
The time dependence of the coupling is taken into account in Eq. 
\r{47} by the stochastic operator $L=L_u+L_v$, where
\begin{equation}
L_u=\nu\left(1+u\frac{\partial}{\partial u}+
\frac{\Omega_0^2}{2}\frac{\partial^2}{\partial u^2}\right).
\label{2}\end{equation}
The random vector $\vec{\Omega}$ has a Gaussian stationary 
distribution $dW(\vec{\Omega})=f(\Omega)d\vec{\Omega}$, where 
$d\vec{\Omega}=dudv$ and
\be
f(\Omega)=\exp(-\Omega^2/\Omega_0^2)/\pi \Omega_0^2.
\e{34}
Assuming that the TLS does not interact with the reservoir at $t<0$,
the initial condition for Eq. \r{47} is 
$G(\vec{\Omega},0)=If(\Omega)$, where $I$ is the unity matrix.

The fully averaged density matrix is given by
\be
\bar{r}(t)=\bar{G}(t)r(0),
\e{66}
where
\be
\bar{G}(t)=\int d\vec{\Omega}G(\vec{\Omega},t).
\e{40}

Alongside with the {\em forward} partial averages $r(\vec{\Omega},t)$ 
and $G(\vec{\Omega},t)$, 
it may be expedient to consider the backward partial averages 
$\tilde{r}(\vec{\Omega}{'},t)$ and $\tilde{G}(\vec{\Omega}{'},t)$ 
related by
\be
\tilde{r}(\vec{\Omega}{'},t)=\tilde{G}(\vec{\Omega}{'},t)r(0),
\e{A6}
where the tilde denotes the average over $\vec{\Omega}(t)$ subject to 
the condition $\vec{\Omega}(0)=\vec{\Omega}{'}$.
As shown in Appendix \ref{A}, the forward and backward partial
averages are related by
\be
G(\vec{\Omega},t)=S^{-2}\tilde{G}{^T}(\Omega,-\phi,t)S^2f(\Omega).
\e{B6}

Henceforth we shall focus on the exact-resonance case,
\be
\Delta_0=0.
\e{46}
Consider a matrix Fourier series 
\be
G(\vec{\Omega},t)=
\sum_{n=-\infty}^{n=\infty}T_n^\dagger G_n(\Omega,t).
\e{67} 
Here $T_n$ is a diagonal matrix, 
\be
T_n=\text{diag}(e^{i(n+1)\phi},e^{in\phi},e^{i(n-1)\phi}),
\e{68}
and $G_n(\Omega,t)=\langle T_nG(\vec{\Omega},t)\rangle_\phi$, where
the average over $\phi$ is denoted by
$\langle\dots\rangle_\phi=(2\pi)^{-1}\int_0^{2\pi}d\phi\dots\ $.
Multiplying Eq. \r{47} by $T_n$ from the left and integrating the 
both sides of the resulting equation with respect to $\phi$ from 0 to 
$2\pi$, one obtains the equations
\be
\dot{G_n}=A(\Omega)G_n+M_nG_n,
\e{69}
where $A(\Omega)=A(\Omega,\phi=0)$ and 
$M_n=\text{diag}(L_{n+1},L_n,L_{n-1})$.
Here $L_k$ is defined so that for an arbitrary function 
$F(\vec{\Omega})$
\be
\langle e^{ik\phi}LF(\vec{\Omega})\rangle_\phi=
L_k\langle e^{ik\phi}F(\vec{\Omega})\rangle_\phi.
\e{57}
On writing $L$ in the polar coordinates \cite{kof01}, one obtains that
\bem{37}
\bea
&&L_k=L_0-k^2\Omega_0^2\nu/2\Omega^2,\label{37a}\\
&&L_0=\frac{\Omega_0^2\nu}{2}\frac{\partial^2}{\partial \Omega^2}+
\left(\nu \Omega+\frac{\Omega_0^2\nu}{2\Omega}\right)
\frac{\partial}{\partial \Omega}+
2\nu.
\ea{37b}
\eml

Taking into account that the only nonzero initial conditions for Eqs. 
\r{69} are $G_{n;-n,-n}(\Omega,0)=f(\Omega)$ ($n=0,\pm 1$), one
obtains from Eqs. \r{69} and \r{67} that
\bea
&&G(\vec{\Omega},t)=\nonumber\\
&&\left(\begin{array}{ccc}
R(\Omega,t)&e^{-i\phi}E(\Omega,t)/2&e^{-2i\phi}P(\Omega,t)\\
-e^{i\phi}Q(\Omega,t)&N(\Omega,t)&e^{-i\phi}Q(\Omega,t)\\
e^{2i\phi}P(\Omega,t)&-e^{i\phi}E(\Omega,t)/2&R(\Omega,t)
\end{array}\right).\nonumber\\
\ea{C2}
The functions entering Eq. \r{C2} satisfy the following sets of
equations,
\be
\dot{N}=-i\Omega E+L_0N,\qquad\dot{E}=-i\Omega N+L_1E
\e{4}
and
\bem{5}
\bea
&\dot{P}=&-(i/2)\Omega Q+L_2P,\ \ \ \dot{Q}=i\Omega(R-P)+L_1Q,
\label{5a}\\
&&\dot{R}=(i/2)\Omega Q+L_0R.
\ea{5b}
\eml
The only nonvanishing initial conditions for Eqs. \r{4} and \r{5} are 
\be
\text{(a)}\ N(\Omega,0)=f(\Omega),\ \ 
\text{(b)}\ R(\Omega,0)=f(\Omega), 
\e{52}
respectively.
The functions $N(\Omega,t)$, $R(\Omega,t)$ and $P(\Omega,t)$ are real,
whereas $E(\Omega,t)$ and $Q(\Omega,t)$ are purely imaginary.
This follows from the fact that, on changing the variables
$E'(\Omega,t)=iE(\Omega,t)$ and
$Q'(\Omega,t)=iQ(\Omega,t)$, Eqs. \r{4} and \r{5} become sets of 
equations with real coefficients and real initial conditions.

Averaging Eq. \r{B1} over $\phi$ yields
\be
r(\Omega,t)=G(\Omega,t)r(0),
\e{70}
where $r(\Omega,t)=\langle r(\vec{\Omega},t)\rangle_\phi$.
As follows from Eq. \r{C2}, 
$G(\Omega,t)\equiv\langle G(\vec{\Omega},t)\rangle_\phi
=\text{diag}(R(\Omega,t),N(\Omega,t),R(\Omega,t))$.
This means that the population relaxation and the coherence 
relaxation proceed independently of each other, 
\bea
&&n(\Omega,t)=N(\Omega,t)n(0),\label{28a}\\
&&\rho_{12(21)}(\Omega,t)=R(\Omega,t)\rho_{12(21)}(0).
\ea{28b} 
Hence, finally,
\be
\bar{n}(t)=N(t)n(0),\quad
\bar{\rho}_{12(21)}(t)=R(t)\rho_{12(21)}(0),
\e{30}
where the population and coherence relaxation functions are obtained 
by 
\bem{31}
\bea
&&N(t)=2\pi\int_0^\infty N(\Omega,t)\Omega d\Omega,\label{31a}\\ 
&&R(t)=2\pi\int_0^\infty R(\Omega,t)\Omega d\Omega.
\ea{31b}
\eml

\section{Limiting cases}
\label{IV}

\paragraph{Zero temperature.}
For the sake of comparison, we mention here
the results obtained for the zero-temperature reservoir with a
Lorentzian spectrum \cite{aga86}.
In this case, for an arbitrary initial state the TLS relaxation
is determined by the equalities $\rho_{22}(t)=N_0(t)\rho_{22}(0)$ and 
$\rho_{12}(t)=R_0(t)\rho_{12}(0)$.
The functions $N_0(t)$ and $R_0(t)$ can decay monotonously or in an
oscillatory fashion, depending on the coupling strength, as described
in Ref. \cite{aga86}.
However, for all values of the coupling strength and detuning one can
show that $N_0(t)=|R_0(t)|^2$, which is the extension of the relation
$T_1=T_2/2$ for a nonexponential relaxation.
The above equality is in a sharp contrast with the present 
infinite-temperature case, as shown below.

\paragraph{Weak coupling.}
Returning to the infinite-temperature case, 
two limiting cases can be readily considered.
For a weak coupling, $\Omega_0\ll\nu$, the average population 
difference and coherence decay exponentially with the decay times 
$T_1=\nu/\Omega_0^2$ and $T_2=2T_1$ respectively \cite{sev89}.

\paragraph{The static limit.}
In the opposite case $\nu\rightarrow 0$ (the static limit) one can 
set $L_n\approx 0$ in Eqs. \r{4} and \r{5}, yielding 
\cite{bur76,bur96,kof78,kni82}
\bem{8}
\bea
&&N_{\rm st}(t)=1-\Omega_0tF(\Omega_0t/2),\label{8a}\\
&&R_{\rm st}(t)=[N_{\rm st}(t)+1]/2,
\ea{8b}
\eml
where $F(z)$ is Dawson's integral \cite{abr64},
\be
F(z)=e^{-z^2}\int_0^zdye^{y^2}.
\e{81}
The function \r{8a} (see Fig. \ref{f1}, the dot-dashed line) 
simplifies in two limits, 
\bem{42}
\bea
&&N_{\rm st}(t)\approx 1-\Omega_0^2t^2/2\ \ (t\ll \Omega_0^{-1}),
 \label{42a}\\
&&N_{\rm st}(t)\approx -2/\Omega_0^2t^2\ \ (t\gg \Omega_0^{-1}).
\ea{42b}
\eml
It vanishes at $t=1.85\Omega_0^{-1}$ and has one minimum equal to 
$-0.285$ at $t=3.00\Omega_0^{-1}$.

As follows from Eq. \r{8b}, 
$\lim_{t\rightarrow\infty}R_{\rm st}(t)=0.5$.
This nonzero limit results from the fact that in the static limit the
component of the TLS pseudospin parallel to the effective field
$\vec{B}=(u,v,0)$ is conserved.
Equation \r{8} implies that for $t\ll \Omega_0^{-1}$ one has 
$N(t)\approx R^2(t)$, as in the weak-coupling case.
In contrast, for $t\gg \Omega_0^{-1}$ the ratio $|N(t)|/R^2(t)$ 
tends to zero.

Note that in the static limit, the relaxation occurs due to the 
statistical spread of the coupling amplitude and, correspondingly, in 
principle it can be reversed by an echo technique.
Irreversible relaxation is obtained only in the presence of temporal 
fluctuations ($\nu\ne 0$).

\paragraph{Short times.}
Consider effects of the temporal fluctuations in the strong-coupling 
regime,
\be
\Omega_0\gg\nu,
\e{50}
for sufficiently short times, i.e., as shown below, for
\be
t^3\ll D^{-1}.
\e{49}
Here $D=\Omega_0^2\nu/2$ is the diffusion coefficient for
$\vec{\Omega}(t)$ [cf. Eq. \r{2}].
For short times, one can use the time-dependent perturbation theory,
as described in Appendix \ref{B}, to calculate the backward partial 
average $\tilde{G}(\vec{\Omega},t)$, which, in turn, yields 
$G(\vec{\Omega},t)$ by Eq. \r{B6}.
Comparing the resulting expression with Eq. \r{C2} yields
\bea
N(\Omega,t)\approx&&f(\Omega)\{a(\Omega,t)\cos \Omega t\nonumber\\
&&+[D(1-\Omega^2t^2)/(2\Omega^3)]\sin\Omega t\},
\ea{C3}
\bea
R(\Omega,t)\approx&&N(\Omega,t)/2+f(\Omega)[1/2
-(Dt/2\Omega^2)(2\nonumber\\
&&+\cos \Omega t)+(3D/2\Omega^3)\sin\Omega t],
\ea{C4}
where $a(\Omega,t)=1-Dt^3/3-Dt/2\Omega^2$. Expressions for 
$E(\Omega,t)$, $P(\Omega,t)$, and $Q(\Omega,t)$ are shown in Appendix
\ref{B}.

\section{Population relaxation}
\label{VI}

First, we discuss the population relaxation.
Equations \r{4} are related to the {\em Schr\"{o}dinger equation for 
some TLS} $\{|a\rangle,|b\rangle\}$ coupled to a resonant chaotic 
field, as follows.
Writing the Hamiltonian of the latter TLS in the form 
\be
H_0=\hbar V_c(t)|b\rangle\langle a|+\text{H.c.},
\e{45}
where $V_c(t)$ is a complex Gaussian-Markovian process,
one obtains the Schr\"{o}dinger equation
\be
\dot{U}_{aa}=-iV_c^*(t)u_{ba},\ \ 
\dot{U}_{ba}=-iV_c(t)u_{aa},
\e{43}
where $U_{aa}(t)$ and $U_{ba}(t)$ are matrix elements of the
evolution operator $\hat{U}(t)$.
If the Hamiltonians \r{35} and \r{45} describe the same interaction,
then $V_c(t)=-\Omega_c(t)/2$ and hence
\be
V=\Omega/2,\ \ V_0=\Omega_0/2,
\e{44}
where $V=|V_c|$ and $V_0^2=\langle V^2\rangle$.

The averaging of the above quantity $\hat{U}(t)$ was discussed in 
Ref. \cite{kof01}.
Writing equations for the partial average quantities
$U_{aa}(\vec{V},t)$ and $U_{ba}(\vec{V},t)$ 
\cite[Eqs. (3.3), (3.7)]{kof01} and
eliminating the field phase, one obtains the equations \cite{note3}
which can be identified with Eqs. \r{4} under the substitutions 
\bea
&&U_{aa}(\vec{V},t)=U_{aa}(V,t)\rightarrow N(\Omega,t),\label{19a}\\
&&U_{ba}(\vec{V},t)e^{-i\phi}\rightarrow E(\Omega,t),\label{19b}\\
&&\text{(a)}\ V\rightarrow \Omega,\ \
\text{(b)}\ V_0\rightarrow \Omega_0.
\ea{19c}
For instance, for $\Omega\ll\Omega_0$, Eq. \r{C3} can be obtained 
from Ref. \cite[Eq. (5.25)]{kof01}, in view of Eqs. \r{49}, \r{19a}, 
and \r{19c}.

\begin{figure}
\vspace*{-3.cm}
\centerline{\psfig{file=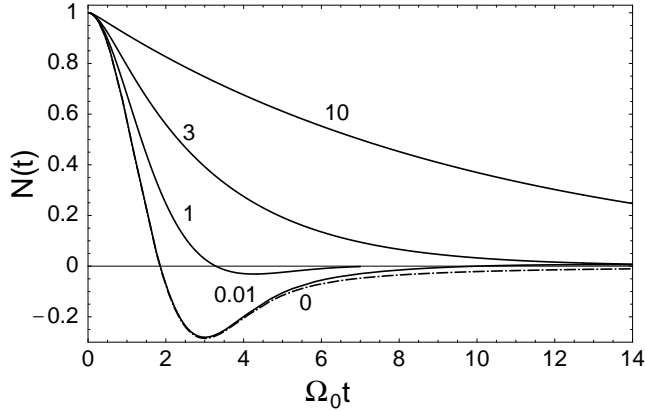,width=3.375in}}
\vspace{-2.5cm}
\caption{The population relaxation function $N(t)$ versus the
dimensionless time $\Omega_0t$ for the values of $\nu/\Omega_0$ shown 
in the plot.
}
\label{f1}\end{figure}

As follows from Eq. \r{19a}, the quantity 
$\bar{U}_{aa}(t)=\int d\vec{V}U_{aa}(V,t)$
equals $N(t)$ under the substitution (\ref{19c}b).
Thus, one can apply directly the results of the comprehensive study 
of $\bar{U}_{aa}(t)$, performed in Ref. \cite{kof01}, to the 
population relaxation function $N(t)$.
Plots of $N(t)$ for different values of $\nu/\Omega_0$ are shown
in Fig. \ref{f1}.

Henceforth, we focus on the strong-coupling regime, Eq. \r{50}.
In this case, as follows from \cite{kof01}, the population relaxation 
is described by 
\be
N(t)=N_{\rm st}(t)J(\alpha t),
\e{7}
where $\alpha=(2D)^{1/3}=(\Omega_0^2\nu)^{1/3}$ and $J(\alpha t)$ is 
a dimensionless function of $\alpha t$, which describes the 
irreversible relaxation.
For short times, the function $N(t)$ is \cite{note4}
\be
J(\alpha t)\approx 1-Dt^3/3.
\e{24}
One can show that the second-order correction to Eq. \r{24} is on
the order of $D^2t^6$, which implies the validity condition \r{50} for
the above short-time results.

\begin{figure}
\vspace*{-3.5cm}
\centerline{\psfig{file=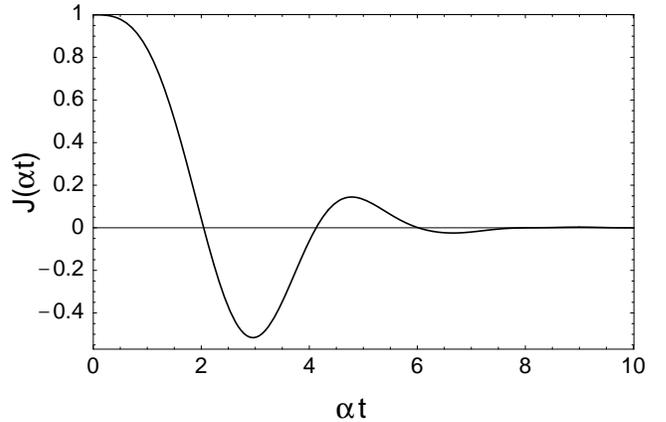,width=3.375in}}
\vspace{-2.5cm}
\caption{The function $J(\alpha t)$ versus the scaled time $\alpha t$.
}
\label{f2}\end{figure}

For $t^3\gg D^{-1}$ $J(\alpha t)$ tends to zero, performing damped 
oscillations (see Fig. \ref{f2}).
Correspondingly, the characteristic rate of the irreversible 
relaxation is on the order of $D^{1/3}\sim\alpha$.

\section{Coherence relaxation}
\label{VII}

Consider the coherence relaxation. 
We shall focus on the strong-interaction regime, Eq. \r{50}.
At short times the coherence relaxation function $R(t)$ is obtained by
inserting Eq. \r{C4} into \r{31b} and performing the integration.
For $t\alt\Omega_0^{-1}$ the $R(t)$ is very close to the static
result, the discrepancy increasing with $t$.
As discussed in Appendix \ref{B}, for 
$\Omega_0^{-2}\ll t^2\ll D^{-2/3}$ 
\be
R(t)\approx 1/2-1/(\Omega_0^2t^2)+\nu t(C_0-{\rm ln}\Omega_0t).
\e{12}
Here $C_0=5/3-\gamma/2=1.38$, where $\gamma$ is the Euler constant
\cite{abr64}.

The coherence relaxation for $t^3\gg D^{-1}$ can be found, as
follows.
Casting Eqs. \r{5} as an equation for the column vector 
\be
q\equiv(q_1,q_0,q_{-1})^T=(P,Q,R)^T
\e{51}
and performing the linear transformation 
$\chi(\Omega,t)=Sq(\Omega,t)$, Eqs. \r{5} become
\be
\dot{\chi}=(i\Omega S_x+M)\chi.
\e{25}
Here $M$ is a diagonal matrix with $M_{mm}=L_{1+m}$.
Note that Eq. \r{25} can be obtained also from partially averaged Eq.
\r{A1}, $\dot{\psi}(\vec{\Omega},t)=i\vec{B}\cdot\vec{S}\psi+L\psi$,
with $\vec{B}=(u,v,0)$, on defining
\be
\chi_m(\Omega,t)=\langle e^{i(1+m)\phi}\psi_m(\vec{\Omega},t)
\rangle_\phi/\psi_{-1}(0)
\e{33}
$(m=1,0,-1)$.
Thus, $\chi(\Omega,t)$ has the meaning of a weighted 
partial average of the wave function for spin 1 in the stochastic 
magnetic field $\vec{B}(t)=(u(t),v(t),0)$.

Next, we invoke the semiclassical dressed-state representation, by
diagonalising the first term on the r.h.s. of Eq. \r{25}.
Since Eq. \r{25} has the form of a (partially averaged)
Schr\"{o}dinger equation, this can be done with the help 
of the standard finite-rotation operator.
Namely, we rotate the $z$ axis around the $y$ axis by $\pi/2$,
$\chi'(\Omega,t)=S_0\chi(\Omega,t)$, where 
$S_{0}=d^{(1)}(\pi/2)$ is defined in \cite{lan77}.
As a result, one obtains
\be
\dot{\chi}{'}=[i\Omega S_z-(D/2\Omega^2)C+L_0]\chi'.
\e{10}
Here 
\be
C=\left(\begin{array}{ccc}
3&-\sqrt{2}&1\\
-\sqrt{2}&4&-\sqrt{2}\\
1&-\sqrt{2}&3\end{array}\right).
\e{11}
The vector $\chi'$ is related to $q$ by $\chi'=S_1q$, where
\be
S_1=S_0S=\left(\begin{array}{ccc}
-1/\sqrt{2}&1/\sqrt{2}&1/\sqrt{2}\\
1&0&1\\
-1/\sqrt{2}&-1/\sqrt{2}&1/\sqrt{2}\end{array}\right).
\e{9}
Correspondingly, the initial condition for Eq. \r{10} is [cf. Eq. 
(\ref{52}b)] $\chi'(\Omega,0)=f(\Omega)\chi^{(0)}$,
where $\chi^{(0)}=(1/\sqrt{2},1,1/\sqrt{2})^T$.

The inverse linear transform $q=S_1^{-1}\chi'$ yields 
\be
R(\Omega,t)={\rm Re}\chi_1'(\Omega,t)/\sqrt{2}+\chi_0'(\Omega,t)/2.
\e{23}
Here we took into account that 
$\chi_1'(\Omega,t)={\chi'}_{-1}^*(\Omega,t)$, which 
follows from the form of Eq. \r{10} and the initial condition for it.
On introducing new variables by 
\be
\chi_m'(\Omega,t)=K_m(\Omega,t)\chi^{(0)}_m,
\e{71} 
Eq. \r{23} becomes
\be
R(\Omega,t)=[{\rm Re}K_1(\Omega,t)+K_0(\Omega,t)]/2.
\e{22}
Note that $K_{-1}(\Omega,t)=K_1^*(\Omega,t)$.
As follows from Eq. \r{71}
\be
K_0(\Omega,0)=K_1(\Omega,0)=f(\Omega).
\e{55}

For $\Omega\gg\alpha$, the nondiagonal terms of the matrix 
coefficient in Eq. \r{10} are much less than the differences of the 
diagonal terms (which are on the order of $2\Omega$) and can be 
neglected in the first (secular) approximation.
This results, in view of Eqs. \r{71}, in the following equations,
\bem{21}
\bea
&&\dot{K}_1=(i\Omega-3D/2\Omega^2+L_0)K_1,\label{21a}\\
&&\dot{K_0}=(-2D/\Omega^2+L_0)K_0,
\ea{21'}
\eml
with the initial conditions \r{55}.

Consider now Eq. \r{21a}.
Note first that in Eqs. \r{21a} and \r{37b} the second terms in the 
parentheses are small as compared to the first terms, respectively, 
and hence can be neglected.
Furthermore, performing the change of variables 
$K_1(\Omega,t)=e^{\nu t}K(\Omega,t)$ Eq. \r{21a} becomes approximately
\be
\dot{K}=(i\Omega+L_\Omega)K,
\e{26}
where $L_\Omega$ is given by Eq. \r{2} with $u\rightarrow\Omega$. 
Equation \r{26} describes the dephasing of a two-level system due to
Gaussian-Markovian frequency fluctuations.
The solution of Eq. \r{26} \cite{rau66} implies that in the
strong-interaction regime \r{50} considered
\be
K_1(\Omega,t)\approx K(\Omega,t)\approx f(\Omega)e^{i\Omega t-Dt^3/3},
\e{61} 
i.e., $K_1(\Omega,t)$ decays with the rate on the order of $\alpha$. 

On the other hand, for $\Omega\alt\alpha$, the equations \r{10} for 
$\chi_m'(\Omega,t)$ are strongly coupled.
An analysis of Eq. \r{10} shows that for $\Omega\alt\alpha$, 
$\chi'(\Omega,t)$ and hence $K_m(\Omega,t)$ [see Eq. \r{71}] 
disappear on the time scale $\alpha^{-1}$.
As a result, in view of Eqs. \r{22} and \r{61}, for $t\gg\alpha^{-1}$ 
\be
R(\Omega,t)\approx\left\{\begin{array}{ll}0,&\Omega\alt\alpha\\ 
K_0(\Omega,t)/2\ \ &\Omega\gg\alpha.\end{array}\right.
\e{56}
Since $\alpha\ll \Omega_0$, the fact that $K_0(\Omega,t)\approx 0$ for
$\Omega\alt\alpha$ can be taken into 
account approximately by the boundary condition to Eq. \r{21'}, 
\be
K_0(0,t)=0.
\e{54}

Equation \r{21'} can be solved by the conjecture
\be
K_0(\Omega,t)=f(\Omega)g(h(t)\Omega),
\e{53} 
where the functions $g(X)$ and $h(t)$ are to be found, using the 
initial and boundary conditions \r{55} and \r{54}.
As shown in Appendix \ref{C}, the solution yields
\be
K_0(\Omega,t)=C_2f(\Omega)\zeta^kM(k,2k+1,-\zeta),
\e{18}
where $k=1/\sqrt{2}$, $\zeta=\Omega^2/\Omega_0^2(e^{2\nu t}-1)$, 
$M()$ is 
the degenerate hypergeometric function \cite{abr64}, and 
\be
C_2=\Gamma(1+k)/\Gamma(1+2k),
\e{58} 
where $\Gamma()$ is the $\Gamma$-function \cite{abr64}.

As follows from Eqs. \r{31b} and \r{56}, 
\be
R(t)\approx K_0(t)/2\ \ (t\gg\alpha^{-1}), 
\e{60}
where 
\be
K_m(t)=2\pi\int_0^\infty K_m(\Omega,t)\Omega d\Omega.
\e{82}
Inserting here Eq. \r{18} and performing the integration with the help
of Ref. \cite[Eq. 7.621.4]{gra65}, one obtains
\be
K_0(t)=C_1e^{-2k\nu t}F(k,k;1+2k;e^{-2\nu t}).
\e{14}
Here $C_1=\Gamma^2(1+k)/\Gamma(1+2k)\approx 0.66$ and $F()$ is 
the hypergeometric function \cite{abr64}.
The function \r{14} monotonously decreases from 1 to 0 with the
average rate on the order of $\nu$.
Equations \r{60} and \r{14} yield the following limits 
\bem{15}
\bea
&&R(t)\approx 1/2+(\nu t/2)({\rm ln}\nu t-C_3)\ \
(\alpha^{-1}\ll t\ll\nu^{-1}),\label{15a}\\
&&R(t)\approx C_1e^{-2k\nu t}\quad(e^{2\nu t}\gg 1).
\ea{15b}
\eml
Here $C_3=1+2k-2\psi(1+k)-2\gamma-{\rm ln}2\approx 0.14$, where 
$\psi()$ is defined in \cite{abr64}.
Equation \r{15a} was obtained with the help of the expansion 
15.3.11 in \cite{abr64} for the hypergeometric function.

Equations \r{12} and \r{15a} describe the behavior of $R(t)$
at $t\ll\nu^{-1}$.
The last term on the right-hand side (rhs) of Eq. \r{12} (which 
arises due to the temporal fluctuations) decreases with the time,
increasing by the magnitude, in contrast to the second term
(pertaining to the static limit).
Note, however, that the last term on the rhs of Eq. \r{12} is much 
less than the second term, i.e., $R(t)\approx R_{\rm st}(t)$ for
$t\ll\alpha^{-1}$.
The magnitudes of the above terms become of the same order
$(\nu/\Omega_0)^{2/3}$ for $t\sim\alpha^{-1}$, where $R(t)$ has a
maximum.
Finally, for 
$t\gg\alpha^{-1}$ $R(t)$ decreases, the coherence dynamics being
determined by the temporal fluctuations [see Eqs. \r{60}, \r{14},
\r{15}].

In the first approximation,
$R(t)$ can be described for all times by the formula
\be
R(t)=R_{\rm st}(t)K_0(t).
\e{16}
Indeed, as follows from  Eq. \r{15a} and the above discussion, for 
$t\alt\alpha^{-1}$ the error in Eq. \r{16} increases
with $t$ from 0 to the value of the order of 
$(\nu/\Omega_0)^{2/3}\ll 1$ at $t\sim\alpha^{-1}$, whereas
for $t\gg\alpha^{-1}$ Eq. \r{16} is close to \r{60}
with the relative error not exceeding by the order of magnitude 
$(\nu/\Omega_0)^{2/3}\ll 1$.

Note that the interpolation formula \r{16} is not unique.
Instead of it, one can use with the same accuracy, e.g., the formula
\be
R(t)=[N_{\rm st}(t)+K_0(t)]/2.
\e{85}
The both formulas provide similar results.

\begin{figure}
\vspace*{-3.cm}
\centerline{\psfig{file=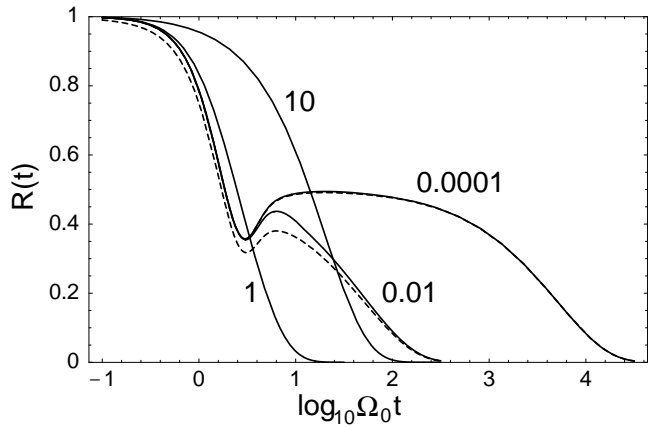,width=3.375in}}
\vspace{-2.2cm}
\caption{The coherence relaxation function $R(t)$ versus 
$\log_{10}\Omega_0t$ for the values of $\nu/\Omega_0$ shown in the
figure.
Solid lines, numerical solution; dashed lines, Eq. \protect\r{16} for
$\nu/\Omega_0=0.01$ and 0.0001.
}
\label{f3}\end{figure}

Figure \ref{f3} demonstrates $R(t)$ for different values of 
$\Omega_0/\nu$.
The solid lines are calculated numerically by inverting the continued 
fraction describing a Fourier transform of $R(t)$ \cite{shi87}, 
whereas the dashed 
lines are the plots of Eq. \r{16} for $\nu/\Omega_0=0.01$ and 0.0001.
As follows from Fig. \ref{f3}, Eq. \r{16} provides a good 
approximation to the exact result in the strong-coupling regime, the
accuracy increasing with the decrease of $\nu/\Omega_0$
(in particular, at $\nu/\Omega_0=0.0001$ the solid and dashed curves
coincide almost completely in Fig. \ref{f3}).

\section{Discussion}
\label{VIII}

As shown above, in the regime of a strong coupling to the reservoir,
$\Omega_0\gg\nu$, the TLS behavior is rather complicated.
The population relaxation is characterized by two time scales.
On the time scale $\Omega_0^{-1}$, where most of the population 
relaxation occurs, the latter is of a reversible character, whereas 
on the time scale $\alpha^{-1}$ the population relaxation becomes
irreversible and proceeds to completion.

The coherence relaxation is even more complicated.
Roughly speaking, half of the coherence decays similarly to the
population relaxation.
However, for $t\gg\alpha^{-1}$, {\em when the population relaxation is
already completed, almost half of the initial coherence still 
survives}.
The decay time of the latter is on the order of $\nu^{-1}$.
This time is greater than the largest population relaxation time 
$\alpha^{-1}$ by the large factor $(\Omega_0/\nu)^{2/3}$.

To explain the above behavior, we write the pseudospin as 
$\vec{s}=\vec{s}_\bot+\vec{s}_\|$, where 
$\vec{s}_\bot=(s_x^\bot,s_y^\bot,n)$ and 
$\vec{s}_\|=(s_x^\|,s_y^\|,0)$ are respectively perpendicular 
and parallel to the initial effective field $\vec{B}(0)=(u',v',0)$.
Here we denoted $\Omega_c(0)=\Omega_c'=u'+iv'=\Omega'e^{i\phi'}$.
The pseudospin component $\vec{s}_\bot$ describes the population 
difference $n$ and the ``out-of-phase' coherence 
$\rho_{21}^\bot=(s_x^\bot+is_y^\bot)/2=ie^{i\phi'}$Im$(\rho_{21}
e^{-i\phi'})$, whereas 
$\vec{s}_\|(t)$ describes the ``in-phase'' coherence 
$\rho_{21}^\|=(s_x^\|+is_y^\|)/2=
e^{i\phi'}\text{Re}(\rho_{21}e^{-i\phi'})$.

Except for the case $\Omega'\alt\alpha$, which has a negligibly small
probability, during the time $\sim\alpha^{-1}$ the direction of 
$\vec{\Omega}(t)$ almost does not change.
As a result, the backward average of $\vec{s}_\bot(t)$ (i.e., the 
average with a given $\Omega_c'$) and hence the backward-averaged
population and out-of-phase
coherence decay at the rate $\alpha$, as in the case of 
collinear (adiabatic) field fluctuations \cite{sli78,kof01,rau66}.
By the same reason, the backward-averaged in-phase coherence 
almost does not decay at $t\alt\alpha^{-1}$.
Actually, it can decay significantly only after the field rotates 
by an angle of the order of $\pi/2$, which, on the average, requires 
the time of the order of the correlation time $\nu^{-1}$.
The averaging of the above quantities over $\Omega_c'$ (i.e., the 
full averaging) does not change the time scales, at which they vanish,
whereas it provides equal contributions from the in- and out-of-phase 
coherences. 
This explains the above fact that the populations and half of the
coherence relax faster than the other half of the coherence.

The above argument can be summarized as follows.
The reservoir considered here is two-dimensional and symmetric with
respect to the rotations in the $xy$ plane (of the pseudospin space).
However, for $t\ll\nu^{-1}$ it behaves effectively as a 
one-dimensional one, since the direction of the effective field 
$\vec{B}(t)$ does not change significantly over short time intervals.
In the strong-coupling regime, this fact results in a substantially
asymmetric behavior of the pseudospin: the component $\vec{s}_\bot(t)$
decays much faster than $\vec{s}_\|(t)$, as discussed above.
In contrast, in the weak-coupling regime, where the decay rates $T_1$ 
and $T_2$ are much longer than the correlation time $\nu^{-1}$ (see 
Sec. \ref{IV}), the short-time behavior of the reservoir is not 
important, and the density matrix elements relax with similar rates.

\section{Pointer states}
\label{IX}

The above results have a bearing on the
the pointer states and related concepts \cite{dal01,fol,zur81,zur93a}.
It is well known (Ref. \cite{zur81} and references therein) that an
important part of any quantum measurement is the reduction of the 
state vector, i.e., the diagonalization of the density matrix of the 
measuring apparatus in a special (pointer) basis, 
the diagonal elements remaining intact.
The apparent contradiction of this process to quantum mechanics, where
any evolution is unitary, has not obtained yet a completely
satisfactory explanation.
The resolution of the above contradiction suggested by Zurek 
\cite{zur81} is that the reduction of the state vector occurs due to
the coupling of the apparatus with the environment.

One can ask which conditions the coupling of the
apparatus (called below the system) and the environment (reservoir)
should satisfy to be capable to produce the state reduction.
There are, at least, two such conditions, as follows.
The first condition requires that the pointer
basis be independent of the state of the reservoir \cite{zur81}.
Another condition stems from the apparent contradiction between the
role, which the reservoir, according to Zurek, plays in obtaining
the state reduction, and the second law of thermodynamics, which
postulates that any interaction with the environment ultimately 
results in the thermal equilibrium of the system, where, of course, 
no information of its initial state is preserved.
The above contradiction can be removed by requiring that in the
pointer basis the relaxation of the off-diagonal elements of the 
density matrix proceed much faster than the relaxation of the diagonal
elements.
Then there exists a nonvanishing time interval, during which a
measurement can be performed.
Hence the second condition states that the system-reservoir 
coupling should be such that the relaxation of the system to 
the equilibrium proceed with significantly different time scales.

Until now, the existence of the pointer states has been verified for
the harmonic oscillator \cite{zur93a} and a collection of many 
two-level systems \cite{fol} weakly interacting with the environment.
As concerns a single TLS, a fast phase relaxation ($T_2\ll T_1$) is 
known to produce the state reduction in the basis of the energy 
eigenstates, as was noted \cite{mil88} in connection with the quantum
Zeno effect \cite{mis77}.

As shown above, in the strong-coupling regime \r{50} the
relaxation of a TLS proceeds with significantly different time
scales, which suggests a possibility of the existence of the pointer 
states.
The discussion in Sec. \ref{VIII} implies that likely candidates for 
the pointer states are the eigenstates $|\psi_{\pm}\rangle$ of the 
initial value of the Hamiltonian, 
$-(\hbar/2)[\Omega_c'|2\rangle\langle 1|+\mbox{H.c.}]$.
One can easily check that
\be
|\psi_{\pm}\rangle=\left(|1\rangle\pm e^{i\phi'}|2\rangle\right)
/\sqrt{2}.
\e{72}

To prove that the states \r{72} are indeed the pointer states, one
should consider the average density matrix conditioned to a fixed
value of the initial field phase $\phi'$,
\be
\tilde{\rho}(\phi',t)=2\pi\int_0^\infty
\tilde{\rho}(\vec{\Omega}{'},t)f(\Omega')\Omega'd\Omega'.
\e{73}
Consider the vector
\be
\tilde{r}(\phi',t)=\tilde{G}(\phi',t)r(0).
\e{75}
The function $\tilde{G}(\phi',t)=2\pi\int_0^\infty
\tilde{G}(\vec{\Omega}{'},t)f(\Omega')\Omega'd\Omega'$ is related to
\be
G(\phi,t)=2\pi\int_0^\infty G(\vec{\Omega},t)\Omega d\Omega 
\e{77}
by [cf. Eq. \r{B6}]
\be
\tilde{G}(\phi',t)=S^{-2}G^T(-\phi',t)S^2.
\e{76}
Combining Eqs. \r{C2}, \r{77} and \r{76} yields
\be
\tilde{G}(\phi',t)=
\left(\begin{array}{ccc}
R(t)&-e^{-i\phi'}Q(t)/2&e^{-2i\phi'}P(t)\\
e^{i\phi'}E(t)&N(t)&-e^{-i\phi'}E(t)\\
e^{2i\phi'}P(t)&e^{i\phi'}Q(t)/2&R(t)
\end{array}\right).
\e{78}
The functions entering Eq. \r{78} are obtained from the functions
appearing in Eq. \r{C2} by the integration, as in Eqs. \r{31}.
One can express $q(t)=(P(t),Q(t), R(t))^T$ [cf. Eq. \r{51}] through 
$K_m(t)$, Eq. \r{82}, using $q(t)=S_1^{-1}\chi'(t)$, where 
$\chi_m'(t)=K_m(t)\chi^{(0)}_m$ [cf. Eq. \r{71}].
This yields
\bem{83}
\bea
&&P(t)=[2K_0(t)-K_1(t)-K_{-1}(t)]/4,\label{83q}\\
&&Q(t)=[K_1(t)-K_{-1}(t)]/\sqrt{2},\label{83b}\\
&&R(t)=[2K_0(t)+K_1(t)+K_{-1}(t)]/4.
\ea{83c}
\eml

Inserting Eq. \r{78} into \r{75} yields $\tilde{r}(\phi',t)$ and hence
$\tilde{\rho}(\phi',t)$ [cf. Eq. \r{32}].
The transformation of the latter to the basis \r{72} is performed by 
$\tilde{\rho}{'}(\phi',t)=\tilde{S}{^\dagger}\tilde{\rho}(\phi',t)
\tilde{S}$, where
\be
\tilde{S}=\frac{1}{\sqrt{2}}\left(\begin{array}{cc}1&1\\
e^{i\phi'}&-e^{i\phi'}\end{array}\right).
\e{79}
Finally, one obtains the following expressions for the components
$\tilde{\rho}_{ij}(\phi',t)$ $(i,j=+,-)$ of
$\tilde{\rho}{'}(\phi',t)$,
\bem{84}
\bea
&&\tilde{\rho}_{++(--)}(\phi',t)=
1/2\pm K_0(t)\text{Re}[\rho_{21}(0)e^{-i\phi'}],\label{84a}\\
&&\tilde{\rho}_{+-}(\phi',t)=\tilde{\rho}_{-+}^*(\phi',t)=
[N(t)+Q(t)]n(0)/2\nonumber\\
&&+i\text{Im}\{[K_1(t)+K_{-1}(t)-2E(t)]\rho_{21}(0)e^{-i\phi'}\}/2.
\ea{84b}
\eml

Equations \r{84} imply that $\tilde{\rho}{'}(\phi',t)$ becomes
diagonal at $t\gg\alpha^{-1}$, since the functions of time appearing 
in rhs of Eq. \r{84b} vanish at such times [see Secs. \ref{VI} and 
\ref{VII} and Eq. \r{83b}].
Taking into account that $K_0(t)\approx 1$ for $t\ll\nu^{-1}$, 
in the time interval
\be
\alpha^{-1}\ll t\ll\nu^{-1}
\e{74}
one obtains that 
\be
\tilde{\rho}{'}(\phi',t)\approx\left(\begin{array}{cc}\rho_{++}(0)&0\\
0&\rho_{--}(0)\end{array}\right),
\e{80}
where $\rho_{++(--)}(0)=1/2\pm\text{Re}[\rho_{21}(0)e^{-i\phi'}]$ are 
the diagonal elements of $\tilde{\rho}{'}(\phi',0)$.

Thus we have proved that in the interval \r{74} the reservoir 
produces the reduction of the state vector in the basis \r{72}.
In contrast to the previous cases described in the 
literature, here the pointer states \r{72} {\em depend on the state of
the reservoir}, due to their dependence 
on the reservoir phase $\phi'$.
Since this violates the above first condition, the strong coupling
considered here is not of a type allowed for the apparatus-environment
interaction.

\section{Conclusion}
\label{X}

Above we presented a comprehensive analysis of the relaxation
of a TLS coupled to an infinite-temperature reservoir.
We showed that in the strong-coupling regime the decoherence can 
proceed much slower than the population relaxation and provided a 
physical interpretation of this effect.
We identified the pointer states, which, in contrast to the previous
findings, appeared to be correlated with the reservoir.

The present results can be checked, e.g., in experiments with
high-$Q$ microwave cavities \cite{rai94,har94}.
Since the theory holds in the infinite-temperature limit, one should
require that, at least, the average number of the photons 
$n_{\rm ph}$ in the resonance mode be large.
Note that for a cavity mode with the frequency 21.5 GHz, used in
experiments in Ref. \cite{rai94}, at the temperature 5 K
$n_{\rm ph}=4.4$.
An increase in the cavity temperature $T$ and/or mode wavelength
$\lambda_c$ can significantly increase the above number, since for
high temperatures ($n_{\rm ph}\gg 1$) $n_{\rm ph}\propto T\lambda_c$.

A more detailed estimation of the experimental conditions would
require a consideration of corrections to the above results 
due to finite, though large, values of $n_{\rm ph}$, which is out of
the scope of the present paper \cite{har84}.
We believe, however, that the main results of this paper, in
particular, those concerning the anomalously slow decoherence and the 
pointer states, will remain valid, at least, qualitatively, also for 
moderately large $n_{\rm ph}$.

\acknowledgments
This work was supported by the Ministry of Absorption.

\appendix

\section{Relation between forward and backward partial averages}
\label{A}

Let us derive the relation between $\tilde{G}(\vec{\Omega}{'},t)$ and 
$G(\vec{\Omega},t)$.
It is convenient to consider first the Green function $U(t)$ of 
Eq. \r{A1} defined by $\psi(t)=U(t)\psi(0)$.
The backward and forward partial averages of $U(t)$ are
denoted by $\tilde{U}(\vec{\Omega}{'},t)$ and $U(\vec{\Omega},t)$.
The latter function satisfies the equation
\be
\dot{U}=[A_1(\vec{\Omega})+L]U
\e{B2}
with the initial condition $U(\vec{\Omega},0)=If(\Omega)$, where 
$A_1(\vec{\Omega})=i\vec{B}\cdot\vec{S}$.
Equations for backward averages were derived in Appendix A of Ref. 
\cite{kof01}.
In particular, Eq. (A11) in Ref. \cite{kof01} implies that 
$\tilde{U}(\vec{\Omega},t)$ obeys the equation
\be
\frac{\partial\tilde{U}{^T}}{\partial t}=
[A_1^T(\vec{\Omega})+L^T]\tilde{U}{^T}
\e{B3}
with $\tilde{U}(\vec{\Omega},0)=I$.

The Gaussian-Markovian random process satisfies the detailed-balance
condition, which implies that Eq. \r{B3} can be recast as
\be
\frac{\partial\tilde{U}{_1^T}}{\partial t}=
[A_1^T(\vec{\Omega})+L]\tilde{U}{_1^T},
\e{B4}
where $\tilde{U}{_1}(\vec{\Omega},t)=\tilde{U}(\vec{\Omega},t)
f(\Omega)$.
Note also that $A_1(\vec{\Omega})\equiv A_1(\Omega,\phi)$, as defined 
above, obeys
$A_1^T(\Omega,\phi)=A_1(\Omega,-\phi)$.
With the account of this, the comparison of Eqs. \r{B2} and \r{B4}
shows that
\be
U(\vec{\Omega},t)=\tilde{U}{^T}(\Omega,-\phi,t)f(\Omega).
\e{B5}
Taking into account that $U(\vec{\Omega},t)=SG(\vec{\Omega},t)
S^{-1}$ and 
$\tilde{U}(\vec{\Omega},t)=S\tilde{G}(\vec{\Omega},t)S^{-1}$, Eq. 
\r{B5}
yields finally Eq. \r{B6}.

\section{Perturbation theory}
\label{B}

Consider the perturbation theory in the strong-coupling regime, Eq. 
\r{50}.
One can write $\vec{\Omega}(t)=\vec{\Omega}'+\vec{W}(t)$, where 
$\vec{\Omega}'\equiv(u',v')=\vec{\Omega}(0)$ and 
$\vec{W}(t)=[W_1(t),W_2(t)]$ obeys $\vec{W}(0)=0$.
Under the unitary transformation $\psi'=S_2\psi$, where 
$S_{2,mm'}=D^{(1)}_{mm'}(\phi',\pi/2,0)=d^{(1)}_{mm'}(\pi/2)
e^{im'\phi'}$ is the finite-rotation matrix \cite{lan77}, 
Eq. \r{A1} becomes
\be
\dot{\psi}{}'=i[\Omega' S_z+A_2(t)]\psi'.
\e{A3}
Here $A_2(t)=-2[W_1'(t)S_z+W_2'(t)S_y]$, where $W_1'(t)$ and 
$W_2'(t)$ are
defined by $W_1'(t)+iW_2'(t)=[W_1(t)+iW_2(t)]e^{-i\phi'}$.

Solving Eq. \r{A3} with the help of the time-dependent perturbation 
theory of the second-order in $A_2(t)$ and averaging the result over 
$\vec{W}{}'(t)$ yields 
\be
\tilde{\psi}{}'(\Omega',t)=\tilde{U}{}'(\Omega',t)\psi'(0),
\e{A5}
where
\bea
&&\tilde{U}{}'(\Omega',t)=U_0'(t)
-i\int_0^tdt_1U_0'(t-t_1)\langle A_2(t_1)\rangle U_0'(t_1)\nonumber\\
&&-\int_0^tdt_1\int_0^{t_1}dt_2U_0'(t-t_1)
\langle A_2(t_1)U_0'(t_1-t_2)A_2(t_2)\rangle \nonumber\\
&&\times U_0'(t_2).
\ea{A2}
Here the tilde and the angular brackets denote the 
average over $\vec{W}{}'(t)$, whereas $U_0'(t)=e^{i\Omega' tS_z}$.

The functions $\Omega'+W_1'(t)$ and $W_2'(t)$ are independent 
identical Gaussian-Markovian 
processes, which equal respectively $\Omega'$ and 0 at $t=0$.
For $t\ll\nu^{-1}$ $W_2'(t)$ can be considered as a diffusion (Wiener)
random process.
The same holds for $\Omega'+W_1'(t)$, unless $\Omega'$ is too large 
in comparison with $\Omega_0$.
As follows from Eq. \r{49}, the approximation 
\r{A2} holds for $t\ll\alpha^{-1}\ll\nu^{-1}$, which means that for
$\Omega'\alt \Omega_0$ $W_1'(t)$ and $W_2'(t)$ in Eq. \r{A2} can be 
considered as independent identical diffusion processes with the 
initial conditions $W_{1,2}'(0)=0$.
Performing the averaging in Eq. \r{A2} with the help of
the relations $\langle W_i'(t_1)\rangle=0$ and
$\langle W_i'(t_1)W_j'(t_2)\rangle=2Dt_2\delta_{ij}$, where 
$\delta_{ij}$ is the Kronecker symbol, yields
\bea
\tilde{U}{}'(\Omega',t)=&&U_0'(t)[1-(Dt^3/3)S_z^2]
-2D\int_0^tdt_1\int_0^{t_1}dt_2t_2\nonumber\\
&&\times U_0'(t-t_1)S_yU_0'(t_1-t_2)S_yU_0'(t_2).
\ea{A4}

Under the inverse linear transformation, 
$\tilde{r}(\vec{\Omega}{'},t)=S_3^{-1}\tilde{\psi}{}'(\Omega',t)$,
where $S_3=S_2S$, Eq. \r{A5} becomes \r{A6}
with
\be
\tilde{G}(\vec{\Omega},t)=S_3^{-1}\tilde{U}{}'(\Omega,t)S_3.
\e{C1} 
From Eqs. \r{A4}, \r{C1}, and \r{B6} one obtains finally 
$G(\vec{\Omega},t)$ and hence Eqs. \r{C3}, \r{C4}, and
\be
E(\Omega,t)\approx-2if(\Omega)\left[\frac{a(\Omega,t)}{2}\sin\Omega t+
\frac{Dt^2}{4\Omega}\cos \Omega t\right],
\e{C5}
\bea
P(\Omega,t)\approx f(\Omega)&&\left[\frac{1}{2}-\frac{1}{2}\left(1-
\frac{Dt^3}{3}-\frac{3Dt}{2\Omega^2}\right)\cos\Omega t\right.
\nonumber\\
&&\left.+\frac{D(1+\Omega^2t^2)}{4\Omega^3}\sin \Omega t
-\frac{Dt}{\Omega^2}\right],
\ea{C6}
\bea
Q(\Omega,t)\approx if(\Omega)&&\left[\left(1-\frac{Dt^3}{3}
-\frac{3Dt}{2\Omega^2}\right)\sin \Omega t\right.\nonumber\\
&&\left.+\frac{D(\Omega^2t^2-4)}{2\Omega^3}\cos \Omega t+
\frac{2D}{\Omega^3}\right].
\ea{C7}

Consider now the calculation of $R(t)$ for 
$\Omega_0^{-2}\ll t^2\ll D^{-2/3}$.
Inserting Eq. \r{C4} into \r{31b} one obtains
\be
R(t)=1/2+N(t)/2+\pi Dt(I_1-2I_2).
\e{C10}
Here $N(t)$ is given by Eqs. \r{7} and \r{24},
\bea
&&I_1=\int_0^\infty \frac{\sin x-x\cos x}{x^2}
f\left(\frac{x}{t}\right)dx\nonumber\\
&&\approx f(0)\int_0^\infty \frac{\sin x-x\cos x}{x^2}dx=f(0),
\ea{C12}
where the approximate equality holds since $t\gg\Omega^{-1}$ and
the latter equality results from Eq. 3.784.4 in Ref. \cite{gra65}, and
\be
I_2=\int_0^\infty \frac{x-\sin x}{x^2}f\left(\frac{x}{t}\right)dx
\equiv J_1+J_2+J_3.
\e{C11}
In Eq. \r{C11} $I_2$ is splitted into three integrals, which are 
calculated for $t\gg\Omega_0^{-1}$ as follows,
\bea
J_1&&=\int_0^1\frac{x-\sin x}{x^2}f\left(\frac{x}{t}\right)dx
\approx f(0)\int_0^1\frac{x-\sin x}{x^2}dx\nonumber\\
&&=f(0)[\gamma+\sin 1-1-\text{Ci}(1)],
\ea{C14}
\be
J_2=-\int_1^\infty\frac{\sin x}{x^2}f\left(\frac{x}{t}\right)dx
\approx f(0)[\text{Ci}(1)-\sin 1],
\e{C15}
and
\bea
J_3&&=\int_1^\infty f\left(\frac{x}{t}\right)\frac{dx}{x}
=\frac{f(0)}{2}E_1\left(\frac{1}{\Omega_0^2t^2}\right)\nonumber\\
&&\approx f(0)(\text{ln}\Omega_0t-\gamma/2).
\ea{C16}
Here $E_1()$ is the integral exponential function and 
Ci() is the integral cosine \cite{abr64}.
The integrals in Eqs. \r{C14} and \r{C15} were calculated with the
help of the Mathematica software \cite{wol96}, whereas the approximate
equality in Eq. \r{C16} results from Ref. \cite[Eq. 5.1.11]{abr64}.
Combining Eqs. \r{C10}-\r{C16} yields Eq. \r{12}.

\section{Solution of Eq. \protect\r{21'}}
\label{C}

Inserting Eq. \r{53} into Eq. \r{21'} and taking into account Eq. 
\r{37b}, one obtains the equation
\be
\frac{\dot{h}X}{h}g'=Dh^2g''+\left(\frac{Dh^2}{X}-\nu X\right)g'
-\frac{2Dh^2}{X^2}g,
\e{C17}
where $X=h(t)\Omega$ and the prime denotes the derivative of $g(X)$
with respect to $X$.
By the assumption, $g(X)$ depends on the time only through $X$, which
means that the coefficients in Eq. \r{C17} should not depend
explicitly on the time.
This is indeed so, if
\be
\dot{h}+\nu h=-Dh^3.
\e{C18}
Then Eq. \r{C17} becomes
\be
g''+(1/X+X)g'-(2/X^2)g=0.
\e{C19}

Equations \r{21'} and \r{55} and the fact that $L_0f(\Omega)=0$ result
in $K_0(\Omega\rightarrow\infty)\rightarrow f(\Omega)$.
The initial and boundary conditions \r{55} and \r{54} and the latter
relation for the function \r{53} yield the following conditions,
\be
\text{(a)}\ g(0)=0,\ \ \text{(b)}\ g(h(0)\Omega)=1,\ \ 
\text{(c)}\ g(\infty)=1.
\e{C20}
The comparison of conditions (\ref{C20}b) and (\ref{C20}c) shows that
$h(0)=\infty$.

Solving Eq. \r{C18} with the latter condition yields 
\be
X=(\Omega/\Omega_0)\sqrt{2/(e^{2\nu t}-1)}.
\e{C21}
The solution of Eq. \r{C19} is obtained by the method in Ref. 
\cite{kam59}.
The substitution $g(X)=(-Y)^kg_1(Y)$, where $k=1/\sqrt{2}$ and 
$Y=-X^2/2$, reduces Eq. \r{C19} to the degenerate hypergeometric
equation \cite[Sec. 2.113]{kam59}
\be
Yg_1''+(2k+1-Y)g_1'-kg_1=0,
\e{C22}
where the prime denotes the derivative with respect to $Y$.
The solution of Eq. \r{C22} yields
\bea
g(X)=&&C_2\zeta^kM(k,1+2k,-\zeta)\nonumber\\
&&+C_2'\zeta^{-k}M(-k,1-2k,-\zeta),
\ea{C23}
where $\zeta=X^2/2$ and $C_2$, $C_2'$ are constants.
The boundary condition (\ref{C20}a) yields $C_2'=0$, whereas the
boundary condition (\ref{C20}c) and the asymptotic formula for $M()$
\cite[Eq. 13.5.1]{abr64} yield Eq. \r{58}.
As a result, Eqs. \r{53} and \r{C23} yield Eq. \r{18}.


\begin{thebibliography}{99}
\bibitem{note3}
Their Laplace transforms in a slightly modified form are given by
Eqs. (B3) in Ref. \protect\cite{kof01}.
\bibitem{note4}
In principle, for short times the function $N(t)$ is obtained by 
inserting Eq. \r{C3} into \r{31a} and performing the integration.
However, noting that for $t\gg\Omega_0^{-1}$ only small 
$\Omega$, $\Omega\ll\Omega_0$, contribute to the integral,
one can use Eq. (5.26) in Ref. \protect\cite{kof01} to obtain Eq.
\r{24}, whereas for $t\alt\Omega_0^{-1}$ $J(\alpha t)$ is very close 
to 1 and one can still use Eq. \r{24}, the error being of the order 
of or less than $\nu/\Omega_0$.
\end{thebibliography}
\end{document}